\documentclass[12pt]{article}
\usepackage{hyperref}
\usepackage[english]{babel}
\title{A non-perturbative argument for the non-abelian Higgs mechanism}

\author{G. De Palma \\ Scuola Normale Superiore and INFN, Sezione di Pisa, Pisa, Italy \\
F. Strocchi \thanks{franco.strocchi@sns.it; fax. +39-050 2214 317} \\ INFN, Sezione di Pisa, Pisa, Italy }
\date{}

\newtheorem{Theorem}{Theorem}[section]

\usepackage{latexsym}
\usepackage[T1]{fontenc}
\usepackage{amsmath}
\usepackage{amssymb}

\def \ra {\rightarrow}

\def \AO {{\cal A}({\cal O})}
\def \AO' {{\cal A}({\cal O}')}

\def \Pf {{\bf Proof.\,\,}}

\def \be {\begin{equation}}
\def \ee {\end{equation}}
\def \ume {{\scriptstyle{\frac{1}{2}}}}
\def \ra {\rightarrow}

\def \eqq {\equiv}

\def \d {{\delta}}
\def \eps {{\varepsilon}}

\def \ph {{\varphi}}

\def \D {{\cal D}}
\def \F {{\cal F}}

\def \L {{\cal L}}

\def \id {{\bf 1 }}

\def \Psio {{\Psi_0}}

\def \dmu {{\partial_\mu}}

\def \d^nu {{\partial^\nu}}
\def \d^la {{\partial^\lambda}}
\def \d^o {{\partial^0}}

\def \xbf {{\bf x}}

\def \Rbf {{\bf R}}

\def\doppio#1{{\rm I}\kern-.1667em{\rm #1}}

\def\Q{\text{Q}\kern-.52em
    \text{\vrule height1.5ex width.5pt depth0pt}\kern.45em}

\def\dZ{{\mathchoice {\hbox{$\Ss\textstyle Z\kern-0.4em Z$}}
{\hbox{$\Ss\textstyle Z\kern-0.4em Z$}} {\hbox{$\Ss\scriptstyle
Z\kern-0.25em Z$}} {\hbox{$\Ss\scriptscriptstyle Z\kern-0.2em
Z$}}}}

\def\dC{{\mathchoice{\hbox{$\rm\textstyle\text{\kern.35em\vrule
   height1.5ex width.5pt depth0pt\kern-.35em C}$}}
{\hbox{$\rm\textstyle\text{\kern.35em\vrule
   height1.5ex width.5pt depth0pt\kern-.35em C}$}}
{\hbox{$\rm\scriptstyle\text{\kern.35em\vrule
   height1.5ex width.3pt depth0pt\kern-.35em C}$}}
{\hbox{$\rm\scriptscriptstyle\text{\kern.35em\vrule
   height1.5ex width.2pt depth0pt\kern-.35em C}$}}}}

\makeatletter

\@addtoreset{equation}{section}

\begin{document}

\maketitle

\begin{abstract}
The evasion of massless Goldstone bosons by the non-abelian Higgs mechanism is proved by a non-perturbative argument in the local BRST gauge

\end{abstract}

\newpage
\section{Introduction}
\vspace{5mm}The extraordinary success of the standard model based on the so-called  Englert--Brout--Higgs--Guralnik--Hagen--Kibble mechanism \cite{1,2,3}, for short the {\em Higgs mechanism}, recently further supported by the detection of a Higgs-like  boson, is one of the biggest achievements of modern theoretical physics.

The present understanding and control of the theory relies on the perturbative expansion and it is in itself an impressive result that one has a renormalized perturbative expansion which incorporates the symmetry breaking condition.

In view of the relevance of such a theory for elementary particle physics, and  for theoretical fundamental physics in general, it is of some interest to have a  general control of the symmetry breaking ansatz at the basis of the perturbative expansion  and a rigorous derivation of the mass generation of the vector bosons with the disappearance  of the massless Goldstone bosons.

The perturbative derivation of such features of the Higgs phenomenon does not exhaust the quest for a more general understanding of the mechanism, since even if the renormalized perturbative expansion is term by term well defined, there is no proof that the series is asymptotic or Borel summable to the exact solution (as unfortunately this is not the case for the $\phi^4$ theory to which the Higgs--Kibble model reduces in the limit of vanishing gauge coupling).

As a matter of fact, whereas the Goldstone theorem has been formulated  with full mathematical  rigor, independently of the perturbative expansion, the same control is not shared  by the standard treatment  of the Higgs mechanism.

Actually, some problems arise in this context: at first sight the Elitzur--De Angelis--De Falco--Guerra (EDDG) theorem \cite{4,5} seems to deny the consistency of the symmetry breaking ansatz (see below), the perturbative expansion is afflicted by the hierarchy problem, the perturbative approach relies on the non-vanishing expectation of the Higgs elementary field and does not cover the case of a symmetry breaking order parameter given by a compound field operator, and the vector boson mass generation is derived by crucially using the  minimal coupling of the Higgs field with the vector boson field.

We briefly recall  the standard discussion of the Higgs mechanism.

The simplest argument about the evasion of the Goldstone theorem goes back to the pioneering papers on the mechanism \cite{1,2,3,6}. It relies on a mean field expansion around the  symmetry breaking vacuum expectation $< \phi >$ of the Higgs field: the quadratic Lagrangian  obtained from  the lowest order of such an expansion does not contain massless scalar bosons and the corresponding vector bosons get a mass from a non-vanishing expectation $< \phi >$. Such a very appealing  argument is usually given without any reference to gauge fixing and, hence, is in conflict with the EDDG theorem, which gives $< \phi > = 0$. As a matter of fact, the analysis of the Higgs mechanism crucially depends on the gauge fixing.

More careful treatments are given in the unitary gauge and in the $\xi$-gauges \cite{7}.

\noindent 1. {\em Unitary gauge}. Its definition relies on a mean field ansatz, since a non-vanishing  vacuum expectation of the Higgs field $< \ph >$ enters in the corresponding gauge fixing,
and therefore is at the basis of the definition itself of the unitary gauge. The Lagrangian is no longer invariant under the global gauge group $G$ and therefore the problem of the evasion of the Goldstone theorem does not even arise. However, such a crucial dependence on a symmetry breaking order parameter requires a self-consistency control of such an ansatz;  it is well known that a mean field ansatz may fail to give the right critical temperature  as well as energy spectrum (as displayed, e.g., by the so-called molecular field approximation of the Heisenberg spin model). Moreover, the perturbative expansion in the unitary gauge is not without problems; in particular, renormalizability fails.

A more convenient class of gauges, which generalize the unitary gauge, is the so-called $\xi$-gauges.

\noindent 2. {\em $\xi$-gauges}. Such gauges generalize the unitary gauge, preserve locality and allow renormalizability, so that a well defined  perturbative series is available. The corresponding gauge fixing
$$ \L_{GF} = - 1/2\xi F^a\,F^a, \,\,\,\,\,F^a \eqq \partial^\mu A_\mu^a - i \xi (t^a)_{n m} < \phi_m > (\phi_n - <\phi_n >)$$
involves the vacuum expectation of the Higgs field (as in the case of the unitary gauge) and the Lagrangian is no longer invariant under the global gauge group $G$.  Thus, the symmetry breaking is not spontaneous and the Goldstone theorem does not apply.

It remains to show the self-consistency of the non-vanishing vacuum expectation of the Higgs field and this can be checked order by order in the renormalized perturbative expansion. Already at the lowest order, i.e. in the expansion of the Lagrangian up to quadratic terms,  the would-be Goldstone bosons have a non-vanishing mass squared proportional to $\xi$ and to the vector boson mass
squared, which is a quadratic function of $< \phi >$; at this order, the non-vanishing  $< \phi >$ is guaranteed by a non-zero minimum of the Higgs potential. Thus, the (perturbative) evasion of the Goldstone massless bosons is rather tricky, since it relies on the assumed non-zero $< \phi >$, i.e. on a mean field ansatz, and one may ask whether a non-perturbative analysis is available.

As is the case for the Goldstone theorem, one would like to have
a more general understanding and control of the Higgs mechanism, based only on the existence   of an order parameter which breaks the gauge symmetry, quite independently of the specific model.

In this note, we present a general non-perturbative argument for the evasion of the Goldstone theorem in the BRST gauge of a (non-abelian) Yang--Mills quantum theory. The theorem proved below generalizes the non-perturbative analysis of the abelian case \cite{8}, by showing that the Golsdtone massless modes which accompany the spontaneous symmetry breaking do not belong to the physical spectrum.

The unphysical nature of the massless modes has been argued within  a perturbative expansion \cite{9}. An attempt toward developing a non-perturbative argument has been proposed  under the crucial  assumptions of the existence of asymptotic limits of the fields of the BRST gauge, the completeness of such asymptotic fields and the existence of poles in the propagators \cite{10}.

Such assumptions are at present not under control and actually questionable; the existence of the asymptotic limits of fields in gauge theory is still an open and debated problem, especially in local gauges, where the field correlation functions define an indefinite inner product space and the space--time translations are not described by unitary (bounded) operators. This problem is still open even in the QED case \cite{11}. Moreover, when a field interacts with massless fields (e.g. with ghost fields) its two-point function does not contain a pole corresponding to a definite mass (the so-called infraparticle spectrum) and in particular the LSZ strategy does not apply. It has been conjectured \cite{11} that a mass pole may, one hopes, show up after a delicate dressing of the fields, but it is not clear how this may allow for  a possible LSZ limit.   For these reasons, the  argument of Ref. \cite{10} may even  look not better founded than the perturbative one.

The aim of this note is to present a relatively simple argument which does not require any ingredient or assumption further than the  BRST gauge.


\section{The Higgs mechanism in Yang--Mills gauge theories}

We recall that the BRST quantization of a Yang--Mills (Y--M) gauge theory is defined by the following gauge fixing \cite{7,12}:
\be{ {\cal L}_{GF} = - \partial^\mu B^a A^a_\mu + \ume \xi B^a B^a - i\,\partial^\mu \bar{c}^a (D_\mu c)^a,}\ee
where $A^a_\mu$ is the gauge vector potential, $a$ runs over gauge group indices,  $c^a$, $\bar{c}^a$ are anticommuting (local) Hermitian fields (the so-called Faddeev--Popov ghosts), $B^a$ is the Nakanishi--Lautrup field and summation over repeated indices is understood.

Since the gauge fixing is invariant under the group $G$ of global gauge transformations, by the Noether theorem there are  corresponding conserved currents $J^a_\mu$.

The Y--M equations of motion read (denoting by $F^a_{\mu \nu}$  the Y--M field strength and by $f^a_{b c}$ the structure constants of the  Lie algebra of the gauge group)
\be{\partial^\nu F^a_{\mu \nu} = J_\mu^a - \dmu B^a - f^a_{ b c} A_\mu^b B^c + i f^a_{b c} \bar{c}^b (D_\mu c)^c}. \label{YMeq}\ee
A very important property of the BRST quantization is that the field algebra $\F$ is local and therefore  one can prove \cite{13} that the conserved  currents $J^a_\mu$ generate the infinitesimal global gauge transformations of the fields:
\be{ i \lim_{R \ra \infty} [ Q_R^a, \,F \,] = \delta^a F, \,\,\,\,\,\forall F \in \F,}\ee
where $Q_R^a$ is a suitably regularized integral of the current density $J_0^a $
\be{ Q^a_R = \int d^4 x J^a_0(\xbf, x_0) f_R(|\xbf|)  \alpha(x_0),}\ee
with $f_R(x) = f(x/R)$, $f \in \D(\Rbf)$, $ f(x) = 1$ for $|x| \leq 1$, $ f(x) = 0$ for $|x| \geq 1 + \eps$, $\alpha \in \D(\Rbf)$, supp \,$\alpha \subseteq [ - \delta, \delta]$, $\tilde{\alpha}(0) = \int d x_0 \,\alpha(x_0) = 1$. Such a local generation of the symmetry does not hold in general in non-local gauges, like the Coulomb gauge of QED \cite{13}.

The vacuum expectation values of the elements of the field algebra  $\F$ define a vector space $\mathcal{D}_0 = \F \Psio$, but the locality of $\F$ requires \cite{14} that the  inner product defined by them
\be{ <  F_1\,\Psio, \,F_2\, \Psio > \eqq < F_1^* \,F_2 >}\label{BRSTcond}\ee
is not semidefinite and not all of the vectors of $\D_0$   describe physical states. The physical vectors $\Psi$ satisfy the BRST subsidiary condition \cite{15,16}
\be{ Q_B \Psi = 0,}\ee
where $Q_B$ is the (nilpotent)  BRST charge.

It is worthwhile to stress that the BRST subsidiary condition (\ref{BRSTcond}) is a necessary condition for the physical vectors. A general argument has been given by S. Weinberg (see Ref. \cite{7}, pp. 32--33).

As in the abelian case, the subsidiary condition is not required for selecting all the vectors with positive inner product (even in the abelian case one may construct vectors with positive inner product which do not satisfy the subsidiary condition \cite{17}). One may reasonably expect that  other additional conditions, beyond positivity,  have to be required for a vector in order that it describes a physical state. The BRST condition is expected (\cite{7}, p. 36 and references therein) to characterize the physical vectors of $\D_0$, but all that is needed for the argument presented in this note is that it is a necessary condition.

The fulfillment  of the BRST condition by the physical vectors may also be argued by noticing that the expectations of the subalgebra of observable fields on the physical states must be the same in any gauge and since the Gauss law holds in the expectations of  the physical subspace of the temporal gauge \cite{18}, this must also be the case for the physical vectors of the BRST gauge (the validity of the Gauss law may be related to the vanishing of the expectation of $(\partial^\nu F^a_{\mu \nu} - J_\mu^a)^2$). By Eq. (\ref{gauss}) below, this is guaranteed by the BRST condition (\ref{BRSTcond}). \goodbreak

\begin{Theorem} [Higgs mechanism] In the BRST gauge of a Y--M theory, if the {\bf global gauge group $G$ is broken} by the vacuum expectation value of an element $F$ of the field algebra $\F$
\be {< \delta^a F > \neq 0,\,\,\,\,\,\,F \in \F,}\label{ordpar}\ee
then the Fourier transform of the two-point function $< J^a_\mu(x) F >$ contains a $\delta(k^2)$, i.e. there are massless Goldstone modes; however, such modes cannot belong to the physical spectrum ({\bf absence of physical Goldstone modes}).
\end{Theorem}
\Pf\, The first part of the Theorem follows from an adaptation \cite{19} of the Kastler, Robinson and Swieca general proof of the Goldstone theorem \cite{20}; the argument exploits  locality and  the Jost--Lehmann--Dyson representation of the local commutator $< [J^a_\mu(x), \,F\,] >$, where $F$ is a generic element of the field algebra $\F$ and need not be an elementary field. In terms of fields which transform as real representations of  the global gauge group $G$, one has, for $R \ra \infty$,
  $$ 2 i \mbox{Im}\, \int d^4 x f_R(\xbf) \alpha(x_0) < J^a_0(\xbf, x_0)\,F > = < [\,Q_R, \,F\,] > \sim $$
\be{ \sim \lambda  \int d^4 x f_R(\xbf) \alpha(x_0)\, \partial_0 D(\xbf, x_0),}\ee
where $D(x)$ is the commutator function of a free massless scalar field and the constant $\lambda$ is different from zero, as a consequence of Eq. (\ref{ordpar}).

\noindent For the proof of the second part of the Theorem one easily sees  that, by using the action of the BRST charge $Q_B$ as the generator of the BRST transformations,  the Y--M equations of motion, Eq. (\ref{YMeq}), may also be written in the following form \cite{21}:
\be{\partial^\nu F^a_{\mu \nu}(x) = J_\mu^a(x) - \{ Q_B, \, (D_\mu \bar{c})^a(x)\,\} \eqq J^a_\mu(x) - \L^a_\mu(x).}\label{gauss}\ee Then, the suitably regularized integral of the zero component of such an equation gives
\be{ < [\, Q^a_R,\, F\,] > = < [ (\partial F^a_0)_R +  (\L^a_0)_R, \,F\,] >.}\ee
By locality \cite{8}, one has that for $R \ra \infty$
\be{ < [ (\partial F^a_0)_R, \,F\,] > \sim 0,}\ee and therefore
\be{ \lambda  \int d^4 x f_R(\xbf) \alpha(x_0)\, \partial_0 D(x) \sim
2 i \mbox{Im}\, \int d^4 x f_R(\xbf) \alpha(x_0) < \L^a_0(x)\,F >.}\ee

Now, there are massless modes in the physical spectrum if physical states contribute as intermediate states to the above two-point function, i.e. if there are physical states $\Psi$ such that
\be{ \lim_{R\ra \infty} < \Psio, \, (\L^a_0)_R \,\Psi > \neq 0.}\ee
However, this is not possible because by the BRST supplementary condition $Q_B \Psi = 0$, $Q_B \Psio = 0$, and therefore
\be{ < \Psio, \,\L^a_0(x)\,\Psi > = < \Psio, (Q_B \, (D_\mu \bar{c})^a(x)  +  (D_\mu \bar{c})^a(x)\,Q_B)\,\Psi > = 0.}\ee

The argument may be further supplemented by the remark that the BRST gauge is Lorentz covariant and therefore only scalar excitations $\Psi$ may have  non-vanishing matrix elements
\be{ \lim_{R \ra \infty} < \Psio, \, \partial^i F_{0 i}(f_R, x_0)\,\Psi >,}\ee
and by covariance and antisymmetry, $< \Psio, F_{0 i} \,\Psi > = 0$. This may be easily seen by considering  physical (improper) eigenstates of the 4-momentum, $\Psi_p$. Then, by Lorentz covariance, $< \Psio, F_{\mu \nu}(0) \,\Psi_p >$ has the form of a sum of second-rank tensors constructed in terms of the 4-vector $p_\mu$, each multiplied by a Lorentz invariant function; $g_{\mu \nu}$ and $p_\mu \,p_\nu$ are the only second-rank tensors and antisymmetry requires the vanishing of their coefficients.
Then
\be{ \lim_{R \ra \infty} < \Psio, \,Q_R\,\Psi > = \lim_{R \ra \infty} < \Psio, \, (\L^a_0)_R\,\Psi >.}\ee
and physical states cannot contribute.

As a technical remark, we note that, as in the abelian case \cite{8}, the discussion of the insertion of a complete set of intermediate states requires us to make reference to a Hilbert--Krein closure ${\cal K}$ of $\D_0$,   in such a way that the (indefinite) inner product $< .,\,. >$ is related to the Hilbert product $( . ,\, . )$ of ${\cal K}$ by a (Hermitian) metric operator $\eta$, $\eta^2 = \id$, $\eta \Psio = \Psio$:
$$ < A \Psio, \, B \Psio > = ( A \Psio, \eta\, B \Psio ), $$
$$ < A\,B > = < \Psio, A\, B\,\Psio > = ( \Psio, \eta A\,B\,\Psio ) = (\Psio, A\,B\,\Psio)$$
and
$$ < Q_R\,F > = \sum_n ( \Psio, Q_R\, \Psi_n ) ( \Psi_n, F\, \Psio) =$$ \be{ \sum_n < \Psio, Q_R \Psi_n > (\Psi_n, \,F \,\Psio).}\ee
 Clearly, the conclusions apply to any Hilbert--Krein closure.


\section{Conclusions}
The above results provide a simple  non-perturbative explanation of the evasion of Goldstone bosons in the breaking of a global gauge symmetry  in the local BRST gauge. The argument, mentioned in the pioneering papers \cite{1,3}, that the Goldstone theorem does not apply in the physical gauges because of the lack of covariance and of locality strictly speaking does not exclude the possibility of the existence of Goldstone bosons associated with the symmetry breaking (even if the proof of the Goldstone theorem does not apply). The theorem discussed above excludes the possibility of that occurrence of physical massless modes in the spectrum of the relevant two-point function associated with the spontaneous symmetry breaking.

\end{document}